\documentclass[%
article%
,a4paper%
,11pt%
,fleqn%
,oneside%
,final%
]{memoir}% 
\usepackage{Huhh_package}% Basic package loading 
\usepackage{Huhh_def_core}% Core definition loading 
\usepackage{Huhh_def_beta}% Teting definition loading 
\setlrmarginsandblock{30mm}{30mm}{*}% left and right
\setulmarginsandblock{30mm}{30mm}{*}%top and bottom
\checkandfixthelayout%

\usepackage{longtable}% 
\usepackage{array}% 
\usepackage[clockwise]{rotating}% 
\usepackage{threeparttable}% 
\usepackage{Huhh_endpack}% 

\newsubfloat{figure}%
\counterwithout{section}{chapter}% Counting by setction 
\numberwithin{corollary}{proposition}% Alternatively, 

\renewcommand{\qedsymbol}{\textcolor{black}{$\blacksquare$}}%
\newcommand{\tunder}[2]{{#1}_{_{#2}}}%
\newcommand{\Hyper}[4]{ \dfrac{\binom{#1}{#2}\binom{#3}{#4}}{\binom{M-1}{G-1}}}%
\def\figscaler{0.48}%

\newcommand{\mytighten}[1]{\myshk#1}%
\def\myshk #1#2{#1\hspace{-.3mm}#2}%

\graphicspath{{./images/}}%

\pretitle{\begin{center}\Large\bfseries}%
\posttitle{\par\end{center}\vskip 0.5em}%
\preauthor{\begin{center}%
\large \lineskip 0.5em%
\begin{tabular}[t]{c}}%
\postauthor{\end{tabular}\par\end{center}}%
\predate{\begin{center}\large}%
\postdate{\par\end{center}}%
\author{%autors in the next line
Huhh, Jun-Sok%
\thanks{%affiliation and contact info. in the nex line 
Department of Economics, Seoul National University. Tel: 82-2-10-4932-9881, E-mail: \href{mailto:anarinsk@gmail.com}{anarinsk@gmail.com}%
}%
}%
\title{%title in the next line 
{Sanctioning by Institution, Skepticism of Punisher\\ and the Evolution of Cooperation}%
\thanks{%notice in the next line 
This version is made only for approved reviewers by the author. If you would refer or have any comment on this article, please contact author via below E-mail.
}%
}%
\myhref{%title info of the PDF output 
Sanctioning by Institution, Skepticism of Punisher and the Evolution of Cooperation}%
{%author info of the PDF output 
HUHH, Jun-Sok}%
\input{FIG_TIKZ.gex}% TikZ figs loading if needed 

\begin{document}%
\begin{titlingpage}%Kill this env if you do not want title page
\maketitle%
{\linespread{1.25}%
\begin{abstract}% 
This article aims to clarify the case and the mechanism where sanction or punishment by institution can deliver the evolution of cooperation. Compared to peer sanctioning, institutional sanctioning may be sensitive to players' attitude toward players who do not pre-commit punishment. Departed from former studies based on the punisher who always acts cooperatively, we assume that the punishing player is skeptical in that she cooperates in proportion to how many same types join in her team. Relying on stochastic adaptive dynamics, we show that institutional sanctioning coupled with skeptical punisher can make cooperation evolve for the case where peer sanctioning may not. 
\vspace{2mm}%
\noindent \keyword{%
public good game, stochastic (adaptive) dynamics, sanction, punishment, institution, fixation probability}%
\noindent \JEL{B52, C73}%
\end{abstract}% 
}
\end{titlingpage}%
\section{Introduction}% 
In theory and practice, sanctioning misbehaviors is core and integrated part in delivering the evolution of cooperation. If defectors are not restricted, they are to increase, which ultimately leads to the demise of cooperation. In preventing defectors from thriving, the role of punishers is considered to be critical, who take their own sacrifice to sanction defectors. However, punishing players tend to be evolutionarily inferior to defectors since the punishing is hard to outperform the punished. For this reason, many studies based on evolutionary game theory and its dynamics has been developed to illustrate how punishers can survive despite its evolutionary disadvantage, and make cooperation evolve in a society. 

This paper touches another aspect of sanctioning, which is not treated frequently in related theories. When we observe some kind of defectors in societal entities, what are our reactions to them? Some may just pass them over, and some may punish them directly; scold them or put some sort of physical actions on them. Another way to punish defection is resorting to some institution such as police or higher ranks that make defection down as a representative of the general good, which we obey. Most of theoretical researches on punishment has conventionally assumed peer sanctioning where players punish defectors directly, which leaves the intriguing issues of institutional sanctioning intact. Based on the methodology of stochastic evolutionary dynamics, this article explores when and how such institutional sanctioning makes cooperation evolve in simple theoretical setting. Our main argument consists of two parts:

\begin{enumerate}[1)]\tightlist
\item Commitment problem is important when institutional sanctioning is applied. If pre-commitment is possible by paying  ex ante some cost of sanctioning, commitment can be done credibly. The public information on commitment level may affect the strategic choice for players, especially punishers who already pay their bill. We show that skepticism of punishers would play a crucial role in making cooperation in a team when it is coupled with institutional sanctioning. This skepticism equips player with the ability to defend itself from unconditional defection and/or to exploit unconditional cooperation. Although our skepticism may not be directly translated into selfishness, the behavior of our skeptical punisher is partly considered to be selfish. In this sense, our model implies that players' selfishness does not always disturb the evolution of cooperation. 
\item In contrast to former studies where sufficient intensity of punishment is assumed for peer sanctioning, our model works well for less intense range of punishment. This implies that the solution by institution can be complementary to that by direct types such as peer punishment. Considering many real-world circumstances that direct and harsh punishment cannot be readily implemented, institutional sanction may fit for this case. 
\end{enumerate}

The organization of the paper is following. \Sref{sec:former} reviews former studies related to our argument. \Sref{sec:setup} elaborates the setup of the model. \Sref{sec:stochastic} provides two logics of evolutionary process that show when and how institutional sanctioning can be effective. \Sref{sec:conclusion} summarizes the gist of the paper, and makes some comments on future researching agendas. 

\section{Related Studies}\label{sec:former}

The first research that inspired this paper is the evolution of cooperation ``via freedom to coercion.'' Based on stochastic evolutionary dynamics, \citet{Hauert_Traulsen_Brandt_Nowak_Sigmund:2007} show that evolutionary dilemma on the origin and the stability of punishment can be solved when lone interaction is introduced. The loner exits her team, and gets a fixed payoff unrelated to others' strategic choice. This loner fixates defectors, and all of the loner are fixated again by the cooperators and the (cooperative) punishers. This evolutionary history ends up with prevailing cooperative state. We would suggest another route for the evolution of cooperation, which works without introducing such lone interaction.

The second is the paradoxical role of ``selfish'' punishment in the evolution of cooperation, which is disposed to punish other defectors even though she acts defectively. \citet{Eldakar_Farrell_Wilson:2007} and \citet{Eldakar_Wilson:2008} assume that strategic choice can be separated from the act of punishment, and show that selfish act can make the evolution of cooperation. Instead of pure selfishness postulated in these studies, we introduce the skepticism that makes players choose their strategy in a team based on the information of punishing commitment. 

The last one is the peculiarity that institutional sanctioning has compared to peer sanctioning \citep{Yamagishi:1986,Gurerk_Irlenbusch_Rockenbach:2006,Kosfeld_Okada_Riedl:2009,Sigmund_Silva_Traulsen_Hauert:2010}. Unlike most of studies that presume peer sanctioning, we introduce institutional sanction that punishment is done via some authority over individual players. With peer sanctioning, the cost of sanctioning can change according to the size of defectors in a team, which may make pre-commitment of sanctioning unbinding. Institutional sanctioning can be credibly committed by paying some costs ex ante before the choice of strategy. \citet{Sigmund_Silva_Traulsen_Hauert:2010} shows that institution may deliver the evolution of cooperation when second-order punishment is to be tackled. By assuming players' utilization of information of commitment, our research investigates more basic and elementary aspect of institutional sanctioning, which makes cooperation evolve in a straight way.

\section{Setups for Sanction by Institution}\label{sec:setup} 

\subsection{Public good game with committing stage}
Our basic framework is a simple game, (linear) public good game (PGG) of $G \geq 3$ size. We consider a well-mixed population of constant size $M \gg G$, and $G$ individuals are randomly selected and offered the option to participate in PGG. Each can contribute for the public good or not; cooperate ($C$) or defect ($D$). For simplicity, players invest a fixed amount $1$ normalized. The contributions of $\tunder{x}{C}$ cooperators in a team are multiplied by $r>2$, and then divided among all $G$ participants. The payoff for each $C$ and $D$ is given by 

\begin{align*}
\begin{cases}
r\dfrac{\tunder{x}{C}}{G}~&\text{for $D$} \\[2.5mm]
r\dfrac{\tunder{x}{C}}{G} -1~&\text{for $C$}.
\end{cases}
\end{align*} 

To integrate commitment on punishment in our model, the game proceeds on three stages at each team level. 

\begin{enumerate}[1)]\tightlist
\item \underline{Committing stage}: Players can pay $\gamma$ to establish an sanctioning institution. If a player pays this cost, she can pre-commit to agree sanctioning defective players. All the participants in a team know this information. 
\item \underline{Contributing stage}: Players participate in PGG described above, and obtain their payoff. 
\item \underline{Sanctioning stage}: Finally, sanctioning mechanism works to punish defective players. Sanctioning technology is implemented in such way that each defector is equally punished by sanctioning institution established in first stage. Net after sanction is the final payoff for each player. 
\end{enumerate}

\subsection{Skepticism of punisher}

The punishing players ($P$) in former studies tend to be naive, who acts cooperatively and punishes others when she observes defectors in her team. We call her {\itshape cooperative punisher} ($\mytighten{CP}$). For them, committing stage is redundant, for they cooperate anyway regardless of information on commitment.

In comparison to $\mytighten{CP}$, we propose a more skeptical type of punisher who chooses her strategy based on information from committing stage. We call her {\itshape skeptical punisher} ($\mytighten{SP}$), who cooperates in proportion to the level of commitment in her team \citep{Rustagi_Engel_Kosfeld:2010}. Although other players' strategies in her team are not known to her, the information known to $\mytighten{SP}$ in a team indicates the intensity of punishment against defection in sanctioning stage. When the punishment is weak, defection can be beneficial for a player since she can exploit unconditional cooperators or defend herself best against unconditional defectors. In sum, $\mytighten{SP}$ is the punisher who is sensitive to the information on commitment in choosing her strategy.

\subsection{Sanctioning mechanism}

The credibility of commitment depends on sanctioning mechanism. To make commitment credible, institutional sanction is to be introduced, which is that each punisher in a team pays a fixed amount to form a local institution to police her team. This institution is used to punish defectors in sanctioning stage. The collected total for punishment in a team, $\gamma {\,} \tunder{x}{P}$, is the total cost of punishment, $\beta \tunder{x}{P}/\tunder{x}{D}$ is imposed on each defector where $\beta>\gamma$ is sanctioning technology, and $\tunder{x}{D}$ and $\tunder{x}{P}$ are numbers of $D$ and $P$ in a team respectively. The $\gamma {\,} \tunder{x}{P}$ is lost for nothing if there be no defectors in a team. On this account, institutional sanctioning can incur social cost when a team consists only of $\mytighten{SP}$.  
%==Reserving 
%%Related studies by evolutionary scientists generally introduce peer sanctioning that each punisher sanctions all of defectors in her team. Peer sanction makes the cost of punishment differentiated by the number of $D$. In our setting, peer sanctioning may not be binding since the cost of sanctioning cannot be known ex ante in committing stage. 

\subsection{Payoffs}

With institutional sanctioning, payoffs ${V}_{i}$ for $i \in \{ C,D, P \}$ have four parts; benefits from cooperation in a team that are equally shared, investment in PGG when the player chooses cooperation, sanction that is inflicted by peer or institution, and cost of sanction incurred if she make commitment for sanctioning institution. After normalizing investment as $1$, payoffs of institutional sanctioning with $\mytighten{CP}$ are given by

\begin{align}\begin{aligned}\label{eqn:payoffCP}
&\tunder{V}{C} (\tunder{x}{C},\tunder{x}{D},\tunder{x}{P}) := r \, \dfrac{\tunder{x}{C}+\tunder{x}{P}}{G} - 1\\
&\tunder{V}{D} (\tunder{x}{C},\tunder{x}{D},\tunder{x}{P}) := r \, \dfrac{\tunder{x}{C}+\tunder{x}{P}}{G} - \beta \, \dfrac{\tunder{x}{P}}{\tunder{x}{D}} \\
&\tunder{V}{P} (\tunder{x}{C},\tunder{x}{D},\tunder{x}{P}) := r \,\dfrac{\tunder{x}{C}+\tunder{x}{P}}{G} - 1 - \gamma,
\end{aligned}\end{align}

\noindent where $r$ is the beneficiary multiplier for PGG. Total population consists of each $\tunder{n}{i}$ for $i \in \{ C,D,P\}$ with $\tunder{n}{C}+\tunder{n}{D}+\tunder{n}{P}=M$.\footnote{For peer sanctioning, $\tunder{V}{D}:=r {\,} \frac{\tunder{x}{C}+\tunder{x}{P}}{G} - \beta {\,} \tunder{x}{P}$, $\tunder{V}{P} := r {\,} \frac{\tunder{x}{C}+\tunder{x}{P}}{G} - 1 - \gamma{\,} \tunder{x}{D}$.} 

In the case of $\mytighten{SP}$, the probability that $\mytighten{SP}$ cooperates is assumed to be simply $\tunder{\delta}{P} (\tunder{x}{P}) := \frac{\tunder{x}{P} - 1}{G-1}$, which is that $\mytighten{SP}$ minds the level of commitment by others in her team to choose her strategy. Payoffs of institutional sanctioning with $\mytighten{SP}$ are given by 

%{\small
\begin{align}\begin{aligned}\label{eqn:payoffSP}
&\tunder{V}{C} (\tunder{x}{C},\tunder{x}{D},\tunder{x}{P}) := r \, \dfrac{\tunder{x}{C} + \tunder{\delta}{P} \tunder{x}{P}}{G} - 1\\
&\tunder{V}{D} (\tunder{x}{C},\tunder{x}{D},\tunder{x}{P}) := r \, \dfrac{\tunder{x}{C} + \tunder{\delta}{P}\tunder{x}{P} }{G} - \beta \, \dfrac{\tunder{x}{P}}{\tunder{x}{D}+(1-\tunder{\delta}{P})\tunder{x}{P}} \\
&\tunder{V}{P} (\tunder{x}{C},\tunder{x}{D},\tunder{x}{P}) := r \,\dfrac{\tunder{x}{C} + \tunder{\delta}{P} \tunder{x}{P}}{G} - \tunder{\delta}{P} \cdot 1 - (1- \tunder{\delta}{P} ) \beta \, \dfrac{\tunder{x}{P}}{\tunder{x}{D}+(1- \tunder{\delta}{P} )\tunder{x}{P}} - \gamma.
\end{aligned}\end{align}
%}

\section{Stochastic Imitation Dynamics}\label{sec:stochastic}

This section discusses two versions of stochastic dynamics where imitation is used as social learning for players. The one is designed for intuitive and analytic purpose; the other is implemented for more general and precise validation of our argument. As is mentioned, only three types, $C$, $D$ and $P$ are to be cast in our scenario. The loner who plays a critical role in \citet{Hauert_Traulsen_Brandt_Nowak_Sigmund:2007} is excluded on purpose to illustrate an alternative route to cooperation without the bypass such as lone interaction. 

\subsection{Simple Imitation Dynamics}

When all of players are composed of one type, this state is absorbing in that imitation or adaptive dynamics would make no change. Namely, each player cannot learn from others in population. We propose a simple imitative dynamics that is heuristic for investigating evolutionary dynamics for our discussion. Particularly, this dynamics is nice to be handled since sampling complication of more generalized processes such as Moran process can be simplified without losing its implications. At first, we put three assumptions to model our simple adaptive process. 

\begin{enumerate}[1)]\tightlist
\item \underline{Adiabatic stochastic process}: When mutation or innovation is introduced, each state can be overturned by these invaders or remain unchanged by the disappearance of them. Resident players who watch invading type are quick to change their strategy if the payoff of invaders is better than theirs. If the resident is better than mutants in payoff, imitation makes mutants disappear. Assuming that such innovations are extremely rare, imitation works much faster than innovation. That is, next mutation always happens after learning process ends up to a homogeneous state. This process can be called ``adiabatic'' since a newly introduced mutation ends up with extinction of this type or with its fixation. These processes can be nicely described by a simple Markov chain with the same number of states of possible players' types \citep{Taylor_Fudenberg_Sasaki_Nowak:2005,Fudenberg_Imhof:2006,Sigmund:2010}. 

\item \underline{Multiple mutants}: Instead of assuming a single invader, we propose that multiple mutants of a type is introduced in a homogeneous state. First, single mutant is not proper to consider more general and complex evolutionary process like Moran process. In Moran process, a mutant that is worse than the resident may not be immediately extinct in the sampling process. This can be partly modeled by approving multiple mutants of a type. Next, mutant type can spring in a group instead of single one if a single mutant has some degree of extra influence over other residents. The size of mutation by type $k$ is denoted by $\mu_k \geq 2$ for $k \in \{ C,D,P \}$. $1 \leq \mu_k \leq G$ is assumed, which is that the size of mutation is not too massive. 

\item \underline{Simultaneous imitation by a universal model}: When homogeneous state is perturbed by the group of a mutant type, our simple imitation process works. After a session of interaction ends, each imitates a universal model who is chosen by its payoff. The choice of the model is based on the size of payoff. If there exists a tie among some of players, one among them are randomly chosen. If there exists a tie among all of players, imitation follows neutral drift where imitation is done by a randomly chosen model. 

\end{enumerate}

Simple Markov transition matrix can be obtained for three stationary states, which can be used in calculating invariant distribution among three states. This transition matrix is given by 

{\small
\begin{align*} %\label{mat:simpleadaptive}
\left( 
\begin{array}{cccc}
1-\tunder{\phi}{\mytighten{DC}} - \tunder{\phi}{\mytighten{PC}} & \tunder{\phi}{\mytighten{CD}} & \tunder{\phi}{\mytighten{CP}} \\
\tunder{\phi}{\mytighten{DC}} & 1-\tunder{\phi}{\mytighten{CD}} -\tunder{\phi}{\mytighten{PD}} & \tunder{\phi}{\mytighten{DP}} \\
\tunder{\phi}{\mytighten{PC}}& \tunder{\phi}{\mytighten{PD}} & 1-\tunder{\phi}{\mytighten{CP}}-\tunder{\phi}{\mytighten{DP}},
\end{array}
\right)
\end{align*}
}

\noindent where $\tunder{\phi}{ij}$ denotes the fixation probability that the absorbing state that consists all of $i$ (all-$i$ state) is overturned to all-$j$ state by invading type $j$ for $i,j \in \{C,D,P\}$. Fixation probabilities can be determined in a very simple way. $\phi_{ij}$ is given by multiplying the probability of $\mu_i$ random mutations and the transition probability by imitation when mutation rate goes to zero \citep{Fudenberg_Imhof:2006}. When the population stay a homogeneous state, this can be perturbed by one type of $\tunder{\mu}{i}$-sized mutants. As is mentioned, after mutants spring, two types of players compare the payoff of their own with that of the other. For our simple imitation dynamics, each fixation probability is given by

\begin{align} \label{eqn:SID}
\tunder{\phi}{ij} :=
\begin{cases}
(\tunder{p}{m})^{\tunder{\mu}{i}} \cdot 1 & \text{if~} \tunder{\pi}{ij}(\tunder{\mu}{i}, M-\tunder{\mu}{i}) > \tunder{\pi}{ji}(\tunder{\mu}{i}, M-\tunder{\mu}{i})\\
(\tunder{p}{m})^{\tunder{\mu}{i}} \cdot 0 & \text{if~} \tunder{\pi}{ij}(\tunder{\mu}{i}, M-\tunder{\mu}{i}) < \tunder{\pi}{ji}(\tunder{\mu}{i}, M-\tunder{\mu}{i})\\
(\tunder{p}{m})^{\tunder{\mu}{i}} \cdot \dfrac{\tunder{\mu}{i}}{M} & \text{if~} \tunder{\pi}{ij}(\tunder{\mu}{i}, M-\tunder{\mu}{i}) = \tunder{\pi}{ji}(\tunder{\mu}{i}, M-\tunder{\mu}{i}),
\end{cases}
\end{align}

\noindent where $\tunder{\pi}{ij}$ is the expected payoff of $i$ against $j$ when $\tunder{\mu}{i}$-sized $i$, $(M-\tunder{\mu}{i})$-sized $j$ with all other types extinct for $i,j \in \{ C,D,P \}$, and $\tunder{p}{m}$ is the probability that a mutant chooses the type $i$ among set of types.\footnote{\aref{proof:proposition:badCP} describes the exact definition of $\tunder{\pi}{ij}$.} For the interaction among $C$, $D$ and $P$, $\tunder{p}{m}$ is given by $1/2$ where a mutant randomly choose one from alternatives in a homogeneous state. Third case of \Meref{eqn:SID} follows the neutral drift where the transition probability by $\tunder{\mu}{i}/M$.\footnote{For the case of neutral drift, a mutant enjoys a same payoff with the resident. This mutant can fixate the population when she is randomly chosen for the model to imitate, which happens by the probability of $\tunder{\mu}{i}/M$.} 

For the class of irreducible transition matrix, invariant distribution among three states can be uniquely given by the eigenvector of the largest eigenvalue, $1$, in our case \citep{Fudenberg_Imhof:2006}. Irreducible transition matrices for this case, however, are not suitable to investigate our problem analytically because the burden of calculation is hard to be handled. Thus, we investigate an extreme case of reducible transition matrix, where the full cooperation is realized. Following lemma shows those cases. 

\begin{lemma}\label{lemma:limitingDist}
Assuming $2<r<G$, the fully cooperative state where $D$ disappears is realized if and only if 
\begin{align*}
\tunder{\phi}{\mytighten{CD}} = 0~(\text{i.e., }\tunder{\phi}{\mytighten{DC}} > 0),~%
\tunder{\phi}{\mytighten{CP}} = 0,~%
\tunder{\phi}{\mytighten{DP}} =0,~%
\tunder{\phi}{\mytighten{PD}} > 0.%
\end{align*}
\end{lemma}

\Mlref{lemma:limitingDist} tells that the full cooperation can be realized only in $P$-all state if $2<r<G$.\footnote{The proof is in \aref{proof:lemma:limitingDist}.} The absorbing state of $P$ cannot be overturned by invasion of $C$ and $D$, and $P$ can fixate $D$-all state. For the case of $r>G$ that the team is excessively productive, the evolution of cooperation may not be serious issue because $C$ can fixate $D$-all state. As a matter of fact, there is no need to introduce $P$ for this case. This is why we restrict our attention to the case of $2<r<G$. Following proposition shows that the evolution of cooperation is not delivered when $\mytighten{CP}$ is introduced with sanctioning institution. 

\begin{proposition}\label{proposition:badCP}
For the interaction among $C$, $D$ and $P$, sanction by institution coupled with $\mytighten{CP}$ cannot deliver the fully cooperative state.\footnote{The proof is in \aref{proof:proposition:badCP}. }
\end{proposition}

As is the case of $\mytighten{CP}$ with sanction by peer \citep{Hauert_Traulsen_Brandt_Nowak_Sigmund:2007}, sanctioning institution cannot make evolution of cooperation for the interaction among $C$, $D$ and $P$. Intuitively, $\mytighten{CP}$ with sanctioning by institution is always invaded by $C$ because benefits between two types are same but $P$ has already paid set-up fee for sanctioning institution. Hence $\tunder{\pi}{\mytighten{CP}} (\tunder{\mu}{C}) > \tunder{\pi}{\mytighten{PC}} (\tunder{\mu}{C})$. 

\begin{corollary}\label{col:badCP}
Institutional sanctioning with $\mytighten{CP}$ cannot make the evolution of cooperation when the loner ($L$) is introduced. 
\end{corollary}

The proof of \Mcref{col:badCP} is trivial. \citet{Brandt_Hauert_Sigmund:2006} shows that $L$ makes evolutionary cycle when it is introduced to the interaction between $C$ and $D$. The interaction among $C$, $D$, $L$ and $P$ makes the evolution of cooperation only if $C$ cannot fixate $P$. For peer sanctioning $C$ and $P$ enjoy the same payoff, which makes neutral drift between two. But, as institutional sanctioning with $\mytighten{CP}$ makes $C$ fixate $P$, the cyclical dynamics among $C$, $D$, $L$ and $P$ emerges, which is similar to the interaction among $C$, $D$ and $L$. 

The result so far implies that institutional sanctioning cannot make its way when $P$ does not utilize information of committing stage. So to speak, $\mytighten{CP}$ casted in most of evolutionary studies cannot validate our sanctioning institution. \Mpref{proposition:badCP} is intriguingly modified when $\mytighten{SP}$ who takes advantage of the information comes in. 

\begin{proposition}\label{proposition:goodSPmu2}
Let us assume that finite $M \gg G$. The fully cooperative state can be delivered by institutional sanction with suspicious punisher ($\mytighten{SP}$) if
\begin{inparaenum}[1)]
\item 2 mutants are introduced, and 
\item the intensity and the cost of punishment are smaller than the properly given.%
\end{inparaenum}\footnote{The proof is in \aref{proof:proposition:goodSPmu2}.}
\end{proposition}

\Mpref{proposition:goodSPmu2} shows that sanctioning institution works nicely when it is coupled with $\mytighten{SP}$. At first, the first condition in the proposition illustrates that invading of $\mytighten{SP}$ into $D$-all state can be done with two mutants springing. Imagine that a single $\mytighten{SP}$ is introduced in $D$-all state. As there is no contribution, $\mytighten{SP}$ always defects, and payoffs from contributing stage between $\mytighten{SP}$ and $D$ are same. However, as $\mytighten{SP}$ pays the setup cost of institution, $\tunder{\pi}{\mytighten{PD}}(\tunder{\mu}{P}=1) < \tunder{\pi}{\mytighten{DP}}(\tunder{\mu}{P}=1)$. When two mutants exist in total population, the chance that those two are teamed up in a same group can open the door for $\mytighten{SP}$ to invade into $D$-all state. 

The second is intriguing since it indicates the condition of $\beta$ and $\gamma$, the effectiveness and the cost of punishment, where institutional sanction works well. Former studies show that peer sanction with exit option can deliver the evolution of cooperation if the punishment is sufficiently effective, and the cost of it is affordable. This is to prevent $D$ from fixating $P$-all state, which can be done only when punishment is sufficiently effective. Our result shows that institutional sanctioning coupled with $\mytighten{SP}$ loses its power when punishment is too harsh. Heuristically, $\mytighten{SP}$ is not unconditional cooperator but opportunistic in the sense that she would take advantage of non-committers. It is noted that institutional sanction is also applied to her if she acts defectively. Too effective punishment may harm $\mytighten{SP}$ seriously, which can hinder $\mytighten{SP}$ from thriving in population. 

We can tell that increasing $\mu$ would make more favorable condition for the fully cooperative state, which is that the proper range of $\beta$ and $\gamma$ expands. Following proposition shows more general results by multiple mutants, $\mu \geq 2$. 

\begin{proposition}\label{proposition:goodSPmuk}
Let us assume that finite $M \gg G$. The fully cooperative state can be delivered by institutional sanction with suspicious punisher ($\mytighten{SP}$) if
\begin{inparaenum}[1)]
\item $\mu \geq 2$ mutants are introduced, and 
\item the intensity and the cost of punishment are smaller than the properly given.\footnote{The Proof is in \aref{proof:proposition:goodSPmuk}.}
\end{inparaenum}
\end{proposition}

\begin{corollary}\label{col:goodSPmuk}
The range for $\beta$ and $\gamma$ for the fully cooperative state
\begin{inparaenum}[1)]
\item increases as $\mu$ gets larger, and
\item decrease as $M$ gets larger.%
\end{inparaenum}
\footnote{The RHS of the proof in \aref{proof:proposition:goodSPmuk} directly proves this corollary.}
\end{corollary}

The second part of \Mcref{col:goodSPmuk} implies that our institutional sanctioning works only when $M$ is not too large. This is interesting compared to the case of peer sanctioning with four types, $C$, $D$, $L$ and $\mytighten{CP}$. For any proper $\beta$ and $\gamma$, according to our simple dynamics, the frequency of cooperative state is given simply by $\frac{4\mu+M}{7\mu+M}$, which converges to $1$ as $M$ get larger.\footnote{Detail derivation is in \aref{case:peerwithexit} } 

As is mentioned, $\mu$-sized mutants of $\mytighten{SP}$ should successfully infiltrate or at least make neutral drift to the homogeneous state of $D$ when the fully cooperative state is realized. This relies on the sampling odd that more than one $\mytighten{SP}$ are selected in a team, which decreases as $M$ grows. If the institution cannot adjust its working on $\beta$ and $\gamma$, well-operated sanctioning institution can turn to be obsolete when the size of total population increases. Limiting case of $M \to \infty$, institutional sanctioning cannot be helpful in the evolution of full cooperation for arbitrarily given $\beta$ and $\gamma$. 

\subsection{Moran Process}

Now, we would extend previous results of institutional sanction with $\mytighten{SP}$ to more general and complex adaptive process. This would show that former simple imitation dynamics can gain wider applicability for the cases where analytic approach cannot be approved. 

The Moran process is a natural way to go in studying stochastic evolutionary dynamics. Moran process is a classical model of population dynamics, which is developed in population genetics, and has been imported to game theory recently. In every time step an individual is randomly chosen for reproduction by its fitness, and makes a single clone that replace a randomly selected other member. The sampling for imitation based on payoffs is continued until the population ends up with a homogeneous state.\footnote{See \citet{Fudenberg_Imhof_Nowak_Taylor:2004} for detail theoretical exposition on Moran Process.} Fixation probabilities under Moran process are given by

\begin{align}\label{eq:fixationprob}
\tunder{\phi}{ij}=\frac{1}{\displaystyle \sum_{k=0}^{M-1} \prod_{m=1}^k \frac{1-s + s \tunder{\pi}{ji}(m)}{1-s+s \tunder{\pi}{ij}(m)}},
\end{align}

\noindent where $\tunder{\pi}{ij}(m)$ is the expected payoff of $i$ type with $j$ type when the population consists of $m$-sized $i$ type and $(M-m)$-sized $j$ type where $i,j \in \{ C,D,P \}$.\footnote{See \aref{proof:proposition:badCP} for exact formulation for $\tunder{\pi}{ij}(m)$. Consult \citet{Traulsen_Hauert:2009} for the friendly derivation of \Meref{eq:fixationprob}.} In Moran process, payoffs are adjusted by $s$ to prevent them from turning negative where $1$ is the baseline payoff, and $s$ is the intensity of selection, which cannot be higher than $1/(1-\min {\tunder{\pi}{ij}})$. As all of fixation probabilities are positive, transition matrix is irreducible, which means that any state of $i$ can be reached by starting $j \neq i$.\footnote{Transition matrix is trivially aperiodic and recurrent, thus invariant distribution exists.}  We make cases of invariant distribution under standard Moran process by parameters properly given.\footnote{Calculating modules are written by MATHEMATICA 8.0 of Wolfram, Inc.}

\begin{claim} \label{claim:Moran}
For Moran process, institutional sanction with suspicious punisher can make the evolution of cooperation for the low intensity and the low cost of punishment properly given.
\end{claim}

\begin{figure}[htf]
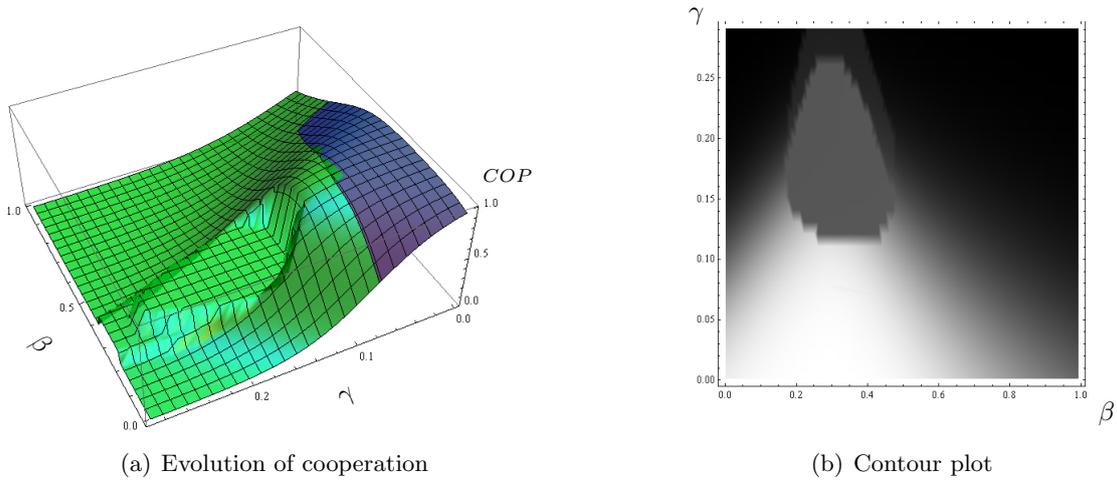
%
\subbottom[Evolution of cooperation]{\standardMoran{0.3}}
\hfill
\subbottom[Contour plot]{\contourSMoran{0.27}}
\caption{The evolution of cooperation by institutional sanctioning with $\mytighten{SP}$ for Moran process. Values are obtained by standard Moran process, and parameters are $M=100$, $G=5$, $r=3$ and $s=0.3$. A computer program generates values by $0.01$ step for $\beta$ and $\gamma$. $C\hspace{-.2mm}O\hspace{-.2mm}P$ is the sum of frequency where population stays in $C$ and $P$. (a) shows the range of $\beta$ and $\gamma$ that makes fully cooperative state. For starker area, $\tunder{\phi}{\mytighten{PD}} \geq 1/M$. (b) draws the contour map, and the gray scale represents the frequency of cooperative state from black ($C\hspace{-.2mm}O\hspace{-.2mm}P=0$) to white ($C\hspace{-.2mm}O\hspace{-.2mm}P=1$). The scale of $\beta$ and $\gamma$ is adjusted for visualization.}
\label{fig:stdMoran}
\end{figure}

\Fref{fig:stdMoran} exemplifies Claim \ref{claim:Moran}. For small $\beta$ and $\gamma$ properly given, the evolution of cooperation is realized by institutional sanctioning coupled with $\mytighten{SP}$. For starker area in (a) of \Fref{fig:stdMoran} where $\tunder{\phi}{\mytighten{PD}} \geq 1/M$, the speed of fixation from $D$ to $\mytighten{SP}$ is fast enough. This ensures stability of cooperative state. Intriguingly, former results by our simple imitation dynamics fairly resemble those by the Moran process that has more general and complex formulation for players' imitation process. 

\begin{figure}[htf]
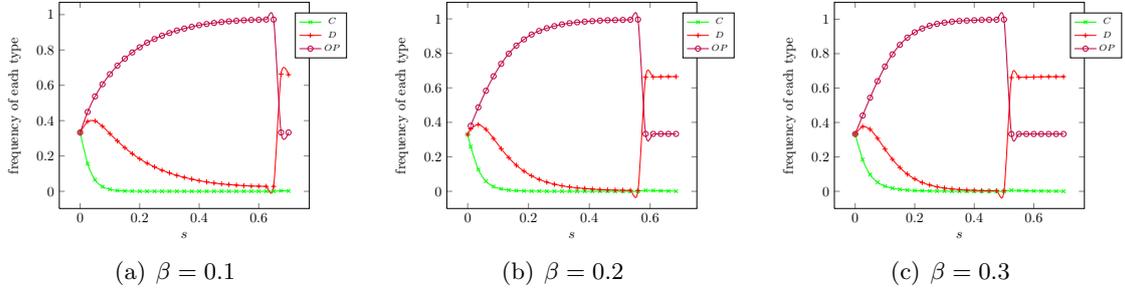
 
\begin{center}
\subbottom[$\beta=0.1$]{\sPlotMoranOne{\figscaler}}
\hfill
\subbottom[$\beta=0.2$]{\sPlotMoranTwo{\figscaler}}
\hfill
\subbottom[$\beta=0.3$]{\sPlotMoranThree{\figscaler}}
\end{center}
\caption{The increase of $s$ on the frequency of each type for standard Moran process. Parameters are  $r=3$, $\gamma=0.05$, $G=5$, $M=100$, and Maximum $s$ is $0.705$. As $\beta$ increases, the level of $s$ that unravel the cooperative state decreases.}
\label{fig:sPlot}
\end{figure}

The intensity of selection, $s$, also affects working of institutional sanction. \Fref{fig:sPlot} illustrates that the effect of $s$ on the frequency of cooperative state changes its direction by some $s$.

\begin{figure}[htf]
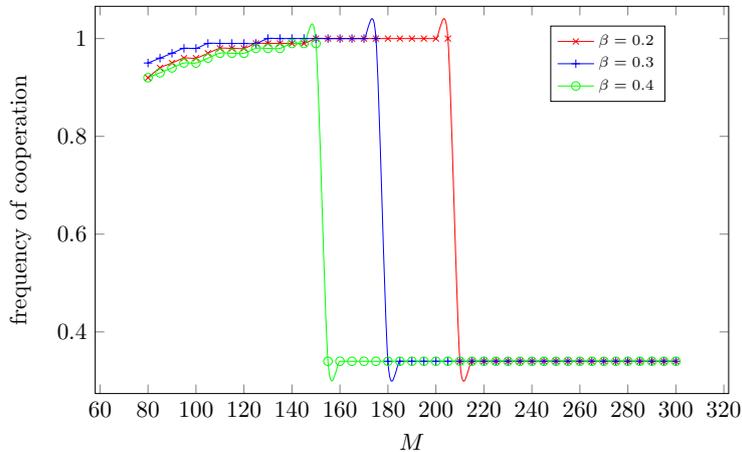
 
\begin{center}
\mPlot{0.8}
\end{center}
\caption{The increase of $M$ on the frequency of cooperation. Parameters are $r=3$, $\gamma=0.05$, $s=0.3$ and $G=5$. When $M$ is larger than a certain level, institutional sanctioning cannot deliver the evolution of cooperation. Moran process makes similar condition to former simple imitation dynamics for $\beta$ and $\gamma$ to diliver the evolution of cooperation. }
\label{fig:mPlot}
\end{figure}

\Mcref{col:goodSPmuk} shows that the size of $M$ has negative effect on the frequency of cooperative state for given $\beta$ and $\gamma$. Moran process for institutional sanctioning replicates this result. \Fref{fig:mPlot} illustrates that the increase of $M$ unravels cooperative state. For given $\beta$ and $\gamma$, as the size of $M$ increases, the frequency that the population stays in cooperative state drops abruptly at a certain level of $M$. This implies that, for Moran process, our institutional sanctioning works effectively within certain level of $M$.

\section{Concluding Remarks}\label{sec:conclusion}
%Sanctioning is critical in restricting defective behaviors in a society. Many studies are concentrated on its evolutionary functioning and origin. 
We've examined the evolution of cooperation in the context of how punishment is done. Considering that most of punishments tend to involve institution, our study fills the gap in researches, which assumes conventional peer sanctioning. Main lessons of this paper are as follows. 

\begin{enumerate}[1)]\tightlist
\item Institution can be justified when the efficacy and the cost of sanctioning both are not too large when they are compared with those of sanctioning by peer. This is reasonable result considering that real-world institution and its sanctioning details. 

\item The work of institution depends on players' skepticism that watch carefully signals from commitment to determine their strategies. Our implication is similar to the insight that is provided by studies of tag-based evolution \citet{Riolo_Cohen_Axelrod:2001}. Ours, however, makes more sense economically than tag-based evolution since issue of institution and commitment is explicitly integrated. 
\end{enumerate}

Sanctioning by institution can provide one of key aspects that the study on the evolution of cooperation should explore. This paper suggests one simple route that institutional sanctioning affects the evolutionary process among different types of players and the evolution of cooperation. Actually, the theoretical setting of this paper is so heuristic that other interesting problems that institutions can embrace are omitted from the consideration. Further studies on many details of institution may enrich the research of evolutionary game theory and its dynamics on the evolution of cooperation. 

\appendix 
\appendix\def\thesection{Appendix \Alph{section}}
%\AppendixTitleToToc
%\renewcommand{\printchapternum}{}
\setcounter{equation}{0}
\numberwithin{equation}{section}
\def\theequation{\Alph{section}.\arabic{equation}}

\section[Appx 1]{Proof of \Mlref{lemma:limitingDist}}\label{proof:lemma:limitingDist}

It is easy to show that $\tunder{\phi}{\mytighten{DC}}>0$ and $\tunder{\phi}{\mytighten{CD}}=0$ are satisfied when $0<r<G$. With this, other fixation probabilities are to be determined that reducible transition matrix ends up with the full cooperative state. Generally, invariant distribution of reducible transition matrix depends on initial condition, but we can find the condition for the fully cooperative state. $\tunder{\phi}{\mytighten{PD}}>0$ and $\tunder{\phi}{\mytighten{DP}}=0$ should hold to prevent $D$ from appearing in invariant distribution, and $\tunder{\phi}{\mytighten{CP}}=0$ hold to prevent the cyclic movement among $C$, $D$ and $P$.\hfill \qedsymbol 

\section[Appx 2]{Proof of \Mpref{proposition:badCP}}\label{proof:proposition:badCP}

When $\mu$ mutants springs in $M$-sized population, they can be grouped as PGG teams in different ways. This process can be calculated by considering hyper-geometric function. Let us denote that $\mu_k = \mu = 2$ for $k \in \{ C,D,P \}$. Expected payoff of a representative type $i$ against type $j$ that is sampled in $G$-sized PGG groups are given by 

\begin{align*}
\tunder{\pi}{ij} (\mu_i)= & \sum_{\tunder{n}{i}=0}^{G-1}\Hyper{\mu-1}{\tunder{n}{i}}{M-\mu}{G-\tunder{n}{i}-1} \tunder{\mathcal{V}}{i}(\tunder{n}{i}) = \sum_{\tunder{n}{i}=0}^{\mu-1}\Hyper{\mu-1}{\tunder{n}{i}}{M-\mu}{G-\tunder{n}{i}-1} \tunder{\mathcal{V}}{i}(\tunder{n}{i}) \\[2mm]
\tunder{\pi}{ji} (\mu_i)= & \sum_{\tunder{n}{j}=0}^{G-1}\Hyper{M-\mu-1}{\tunder{n}{j}}{\mu}{G-\tunder{n}{j}-1} \tunder{\mathcal{V}}{j}(\tunder{n}{j}) =\sum_{\tunder{n}{j}=G-\mu-1}^{G-1}\Hyper{M-\mu-1}{\tunder{n}{j}}{\mu}{G-\tunder{n}{j}-1} \tunder{\mathcal{V}}{j}(\tunder{n}{j}),
\end{align*}

\noindent where $\tunder{\mathcal{V}}{i}(\tunder{n}{i}):=V(\tunder{n}{i}+1,G-\tunder{n}{i}-1,\tunder{n}{k}=0)$, $\tunder{n}{i}$ is the number of type $i$ excluding type-$i$ one considered, the number of type $j$ is $G-\tunder{n}{i}-1$, and the number of $k \neq i,j$ type, $\tunder{n}{k}$, is $0$ for $i,j \in \{ C,D,P \}$.

The proposition can be easily proved by showing that the first part of the lemma cannot be satisfied. It is sufficient to show that $\tunder{\phi}{\mytighten{CP}}=0$ cannot hold. As is assumed, $P$ is $\mytighten{CP}$. The payoff of invading $\mu$-sized $C$ against resident $P$ 
and that of resident $P$ against invading $C$ are given by 

\begin{align*}
&\tunder{\pi}{\mytighten{CP}}(\tunder{\mu}{c}) = r + 1 \\
&\tunder{\pi}{\mytighten{PC}}(\tunder{\mu}{c}) = r + 1-\gamma. 
\end{align*}

When $\gamma>0$, $\tunder{\pi}{\mytighten{CP}}(\tunder{\mu}{C})>\tunder{\pi}{\mytighten{PC}}(\tunder{\mu}{C})$ for any positive $\tunder{\mu}{C}$. Thus, all the resident $P$ turn into $C$ by imitation. Fixation probability of $C$ in $P$-all state, $\tunder{\phi}{\mytighten{CP}}$, is given by $(1/2)^\mu \cdot 1$. $\tunder{\phi}{\mytighten{CP}}$ is always positive, which violates \Mlref{lemma:limitingDist}.\hfill \qedsymbol 

\section[Appx 3]{Proof of \Mpref{proposition:goodSPmu2}}\label{proof:proposition:goodSPmu2}

We check the first conditions of \Mlref{lemma:limitingDist}. For $\tunder{\phi}{\mytighten{CD}}=0$, some calculation can show that $\tunder{\pi}{\mytighten{CD}} (\tunder{\mu}{C}) - \tunder{\pi}{\mytighten{DC}} (\tunder{\mu}{C})$ is given by $\frac{r (M-G)}{G (M-1)}-1$, which is negative for $r<G$. Thus, this condition is easily satisfied for any positive $\tunder{\mu}{C}$. Other three conditions of the first line of \Mlref{lemma:limitingDist} are simplifed and arranged by condition of $\beta$ and $\gamma$. All of three conditions are linear with respect to $r$, and are given by

\begin{align}
\begin{aligned}
&\tunder{\pi}{\mytighten{PD}} (\tunder{\mu}{P}) - \tunder{\pi}{\mytighten{DP}}(\tunder{\mu}{P}) = \tunder{A}{1} + \tunder{B}{1}{\,} r \\ 
&\tunder{\pi}{\mytighten{CP}} (\tunder{\mu}{C}) - \tunder{\pi}{\mytighten{PC}}(\tunder{\mu}{C}) = \tunder{A}{2} + \tunder{B}{2}{\,} r \\ 
&\tunder{\pi}{\mytighten{DP}} (\tunder{\mu}{D}) - \tunder{\pi}{\mytighten{PD}}(\tunder{\mu}{D}) = \tunder{A}{3} + \tunder{B}{3}{\,} r 
\end{aligned}
\end{align} 

It is easy to show that $\tunder{B}{1}>0$, $\tunder{B}{2}<0$, and $\tunder{B}{3}<0$. To sufficiently satisfy other three conditions in the lemma,

\begin{align} %\label{eq:con:propermu2}
%\begin{aligned}
& 0 \leq -\dfrac{\tunder{A}{1}}{\tunder{B}{1}}(\equiv r_p) \leq r \label{eq:rp2} \\ 
& \dfrac{\tunder{A}{2}}{\tunder{B}{2}} < \dfrac{\tunder{A}{1}}{\tunder{B}{1}} \label{eq:rc2} \\ 
& \dfrac{\tunder{A}{3}}{\tunder{B}{3}} < \dfrac{\tunder{A}{1}}{\tunder{B}{1}}\label{eq:rd2}.
%\end{aligned}
\end{align} 

\Meref{eq:rp2} is that $r$ is to be properly defined if $P$ can fixate $D$. \Meref{eq:rc2} and \Meref{eq:rd2} mean that any of $C$ and $D$ cannot fixate $P$ for $r > \tunder{r}{p}$. With tedious arithmetics, these conditions boil down to

%{\tiny
\begin{align}\label{eq:con:properbetagamma2}
\gamma + \left(\frac{(G+1) M^2-G (G+5) M+2 G (G+1)}{G (G+1) (M-2) (M-1)}\right) \beta \leq \frac{M - 2G + 2}{M^2-3 M+2},
\end{align} 
%}

\noindent where $2<r<G$ for $G \ll M$, and \Meref{eq:con:properbetagamma2} defines the proper range for $\beta$ and $\gamma$ in the proposition. \hfill \qedsymbol

\section[Appx 4]{Proof of \Mpref{proposition:goodSPmuk}}\label{proof:proposition:goodSPmuk}

The technique to prove this proposition is identical to \Mpref{proposition:goodSPmu2}, but the calculation cannot be done for general $\mu$. For our purpose, it is sufficient to show that the institutional sanctioning works with $\mytighten{SP}$ for some proper $\beta$ and $\gamma$. Our strategy to prove is following. If institutional sanctioning with $\mytighten{SP}$ works for less punishment upon $D$ and more upon $P$, it still works for the case in the proposition. The less and more system can be expressed neatly. We replace $\tunder{V}{D}$ and $\tunder{V}{P}$ in \Meref{eqn:payoffSP} with 

\begin{align} \label{eqn:modV}
\begin{aligned}
&\tunder{V}{D} (\tunder{x}{C},\tunder{x}{D},\tunder{x}{P}) := r \, \dfrac{\tunder{x}{C} + \delta_p }{G} - \beta \, \dfrac{\tunder{x}{P}}{G} \\
&\tunder{V}{P} (\tunder{x}{C},\tunder{x}{D},\tunder{x}{P}) := r \,\dfrac{\tunder{x}{C} + \delta_p }{G} - \delta_p \cdot 1 - \beta (1- \delta_p ) \, \dfrac{G}{\tunder{x}{D}} - \gamma.
\end{aligned}
\end{align}

The punishment term for $D$ in \Meref{eqn:modV} is smaller than that in \Meref{eqn:payoffSP}, that for $P$ is larger. If sanctioning works with less effective punishment, this must also hold for more effective punishment. 
%This comes from the fact that less effective punishment is bad in regulating defectors in the population. 
Thus, the range of $\beta$ and $\gamma$ with \Meref{eqn:modV} to realize the full cooperation is a sufficient condition for that with \Meref{eqn:payoffSP}. The actual proof proceeds similarly on \aref{proof:proposition:goodSPmu2}, and the condition for $\beta$ and $\gamma$ with $\mu$ is given by 

%{\tiny
\begin{align}\label{eq:con:properbetagammak}
\gamma + \left(\frac{G}{G-1} - \frac{G(G-1)\mu}{G(M-1)} \right) \beta \leq 
\frac{(M-2G+2)(\mu -1)}{(M-2)(M -1)},
\end{align} 
%}

\noindent where $2<r<G$ for $G \ll M$. %\Meref{eq:con:properbetagammak} implies that the range of $\beta$ and $\gamma$ get larger as $\mu$ get higher. 
\hfill \qedsymbol 

\section[Appx 5]{Frequency of Cooperative State with Peer Sanctioning Applied among $C$, $D$, $L$ and $\mytighten{CP}$} \label{case:peerwithexit}

For interactions among $C$, $D$, $L$ and $\mytighten{CP}$, the sufficient level of cooperative state can be made when following fixations are realized: 
\begin{inparaenum}[1)]
\item $D$ fixates $C$,
\item $L$ fixates $D$,
\item $C$ and $\mytighten{CP}$ fixates $L$, and 
\item $C$ and $\mytighten{CP}$ are drifted neutrally, 
\end{inparaenum}
and any other fixating relations are not possible for proper case of the evolution of cooperation.

The first is given by $2<r<G$. The second is easily justified when the payoff of exit option is higher than $0$ that is the payoff of $D$-homogeneous state. The third can be understood when the payoff of full cooperative state is higher than that of exit option. The last one merits discussion. In the evolutionary dynamics of ``via freedom to coercion'', the population stays sufficiently long at cooperative state. This can be ensured by neutral drift between $C$ and $\mytighten{CP}$. This drift is successfully defended from $D$, for $P$ still works in the population. When population is fixated by $C$, $D$ can prospers, but which state is quickly taken over by $L$ again. When above four fixating relations are satisfied, $\mu$-sized imitation dynamics makes the transition matrix which is given by

{\small
\begin{align*}
\left( 
\begin{array}{cccc}
1- \left(\frac{1}{3}\right)^\mu - \left(\frac{1}{3}\right)^\mu \frac{\mu}{M} & 0 & \left(\frac{1}{3}\right)^\mu & \left(\frac{1}{3}\right)^\mu \frac{\mu}{M} \\[2mm]
\left(\frac{1}{3}\right)^\mu & 1- \left(\frac{1}{3}\right)^\mu& 0 & 0 \\[2mm]
0 & \left(\frac{1}{3}\right)^\mu & 1- 2 \left(\frac{1}{3}\right)^\mu & 0 \\[2mm]
\left(\frac{1}{3}\right)^\mu \frac{\mu}{M} & 0 & \left(\frac{1}{3}\right)^\mu & 1- \left(\frac{1}{3}\right)^\mu \frac{\mu}{M} 
\end{array}
\right).
\end{align*}
}

This matrix is irreducible, and its unique invariant distribution is given by the normalized eigenvector of eigenvalue $1$. The frequency of cooperative state yields $\frac{4\mu+M}{7\mu+M}$.

%\newpage
\bibliographystyle{aer} % change style of bibtex if necessary 
\bibliography{anarinsk-1stnuke}

\end{document}